\documentclass[10pt]{article}

\usepackage[T1]{fontenc}
\usepackage[utf8]{inputenc}
\usepackage[english]{babel}
\usepackage{amsmath,latexsym,amsfonts,amsthm,bm,mathtools}
\usepackage{graphicx}
\usepackage[top=1.5in, bottom=1.5in, left=1.25in, right=1.25in]{geometry}
\usepackage{booktabs,url,tikz}
\usetikzlibrary{positioning,arrows.meta}
\usepackage[makeroom]{cancel}
\usepackage{dsfont}
\usepackage{relsize}
\usepackage{multirow}

\usepackage{algorithmicx}
\usepackage{algpseudocode}
\usepackage{algorithm}
\newcommand{\Prob}[1]{\mathbb{P}(#1)}

\newcommand{\ProbEst}[1]{\widehat{\mathbb{P}}(#1)}

\newcommand{\fuzzyset}[1]{\xi_{\widetilde{#1}}}
\newcommand{\fuzzysetij}[2]{\xi_{\widetilde{#1}_{#2}}}

\DeclareMathAlphabet{\mathcall}{OMS}{zplm}{m}{n}

\title{Jointly modeling rating responses and times with \\fuzzy numbers: An application to psychometric data}
\author{Niccol\`o Cao$^{\ast}$, Antonio Calcagn\`{i} \\\\
		\footnotesize{\sl University of Padova}\\
		\footnotesize{$\ast$ E-mail: niccolo.cao.research@gmail.com}
	}

\date{}

\begin{document}

\maketitle

\begin{abstract}
In several research areas, ratings data and response times have been successfully used to unfold the stage-wise process through which human raters provide their responses to questionnaires and social surveys. A limitation of the standard approach to analyze this type of data is that it requires the use of independent statistical models. Although this provides an effective way to simplify the data analysis, it could potentially involve difficulties with regards to statistical inference and interpretation. In this sense, a joint analysis could be more effective. In this research article, we describe a way to jointly analyze ratings and response times by means of fuzzy numbers. A probabilistic tree model framework has been adopted to fuzzify ratings data and four-parameter triangular fuzzy numbers have been used in order to integrate crisp responses and times. Finally, a real case study on psychometric data is discussed in order to illustrate the proposed methodology. Overall, we provide initial findings to the problem of using fuzzy numbers as abstract models for representing ratings data with additional information (i.e., response times). The results indicate that using fuzzy numbers lead to theoretically sound and more parsimonious data analysis methods, which limit some statistical issues that may occur with standard data analysis procedures.\\

\end{abstract}

\vspace{1.25cm}

\section{Introduction}\label{sec1}

Rating scales and questionnaires are widespread in behavioral and social sciences and are especially useful in collecting human opinions, attitudes, and socio-demographic information. A typical rating task entails a multicomponential sequence of cognitive tasks which drive raters to provide their response by selecting one of the possible response categories {\cite{schwarz2001asking}}. It is well-accepted that mining the raters' response process can provide new insights into the mechanisms underlying rating choices \cite{tourangeau2000psychology,rosenbaum2011making}. To this end, fuzzy set theory has been widely applied in modeling the non-random and subjective components of the rating response (for a recent review, see \cite{calcagni2021modeling}). By and large, two general approaches can be recognized in the fuzzy rating literature, namely fuzzy direct scales and fuzzy conversion scales. While the former asks raters to provide their response by means of a stage-wise methodology, which is in turn supposed to elicit the subjective components of the rating response (e.g., see \cite{hesketh1988application,de2014fuzzy,calcagni2014dynamic}), the latter aims at turning standard rating data into fuzzy numbers by means of expert-based or statistical-based procedures (e.g., see \cite{li2016indirect,lalla2005ordinal,yu2007fuzzy}). Despite the differences on the way of mapping fuzzy numbers to the rating process, both provide a valuable strategy to avoid the loss of subjective information entailed by standard rating response formats {\cite{costas1994application,chen2015measuring}}.

Recently, a novel fuzzy conversion procedure (fIRTree) has been proposed with the aim of quantifying a particular component of the rating process, namely the rater's decision uncertainty \cite{calcagni2021firtree,calcagni2022psychometric}. The fIRTree procedure is based upon the use of the Item Response Theory-based trees (IRTree), which model the stage-wise rating processes in terms of linear or nested statistical trees \cite{Boeck_2012}. In particular, once an IRTree model has been fit to crisp rating data, fIRTree uses the estimated IRT parameters to map parametric fuzzy numbers to crisp ratings data. In addition, when the four-parameter triangular fuzzy numbers are used \cite{dombi2018flexible}, fIRTree allows for integrating rating responses and response times into a unique parametric representation. In doing so, fIRTree provides a more flexible and parsimonious representation of uncertainty in ratings data.

In this paper, we describe how parametric fuzzy numbers can be easily used to integrate multiple sources of rating information, such as rating responses and response times. The aim is twofold. First, we show how the four-parameter triangular fuzzy numbers~\cite{dombi2018flexible} can be used to jointly model fIRTree-based raters' responses and time in a unique formal representation \cite{calcagni2021firtree}. The importance of response times in quantifying characteristics of the rating process (e.g., item/question's difficulty) has been widely established in the psychometric literature \cite{kyllonen2016use}. Second, we describe a way to analyze and make inference on this type of data by means of an integrated fuzzy statistical framework. To this end, we adopt an epistemic-based fuzzy normal linear model with crisp predictors, where the problem of point estimation for the unknown parameters is addressed using the minimum inaccuracy principle \cite{corral1984minimun}. 
{Fuzzy regression models are widespread methods to analyze fuzzy data \cite{kruse2016fuzzy,couso2019fuzzy,chukhrova2019fuzzy}. According to the epistemic interpretation of a fuzzy set \cite{couso2014statistical}, fuzzy data are affected by two types of imprecision: stochastic imprecision--which is related to the probabilistic model underlying the observations---and possibilistic imprecision---which is in turn associated with an incomplete knowledge of the originally crisp observations \cite{gebhardt1998fuzzy,denoeux2011maximum}.} {Finally, a real case study based on psychometric data is used in order to highlight the features of the proposed approach with regards to more traditional data analysis~procedures.}

The remainder of this paper is structured as follows. Section~\ref{sec2} presents an overview of fuzzy numbers, fIRTree models, and the fuzzy normal linear model with crisp predictors. Section~\ref{sec3} illustrates an application of the new methodology on clinical questionnaire data along with a comparison with standard data analyses. Finally, Section~\ref{sec4} provides final remarks on the current findings.

\section{Methodology}\label{sec2}

\subsection{Fuzzy numbers}\label{sec2_1}

\begin{figure}[H]
	\centering
	\resizebox{7cm}{!}{
		\input{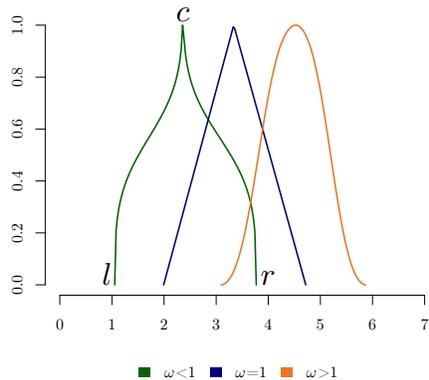}
	}
	\caption{Exemplification of four-parameter triangular fuzzy numbers represented in terms of the $\omega$ parameter. Note that $c$, $l$, $r$ are the center, left and right bounds respectively.\label{fig1}}
\end{figure}

A fuzzy subset $\tilde{A}$ of a universal set $\mathcall A$ is defined by its membership function $\fuzzyset{A}:\mathcall{A}\to [0,1]$. It can be described as a collection of crisp subsets called $\alpha$-sets, i.e., $\tilde{A}_\alpha = \{y \in \mathcall A: ~\fuzzyset{A}(y) > \alpha \}$ with $\alpha \in (0,1]$. If the $\alpha$-sets of $\tilde{A}$ are all convex sets, then $\tilde{A}$ is a convex fuzzy set. The support of $\tilde{A}$ is ${A}_{0} = \{y \in \mathcall A: ~\fuzzyset{A}(y) > 0 \}$ and the core is the set of all its maximal points ${A}_{c} = \{y \in \mathcall A: ~\fuzzyset{A}(y) = \max_{y \in \mathcall A}~ \fuzzyset{A}(y) \}$. In the case where $\max_{y\in \mathcall A} \fuzzyset{A}(y) = 1$, then $\tilde{A}$ is a {normal} fuzzy set. If $\tilde{A}$ is a normal and convex subset of $\mathbb R$, then $\tilde{A}$ is a fuzzy number. The class of all normal fuzzy numbers is denoted by $\mathcall F(\mathbb{R})$. Fuzzy numbers can be represented using parametric models that are indexed by some scalars, such as $c$ (mode) and $s$ (spread or precision). A particular type of parametric fuzzy number is the so-called four-parameters triangular fuzzy number \cite{dombi2018flexible}:
\begin{equation}
	\xi_{\widetilde{A}}(y;c,l,r,\omega) = \bigg(1+\bigg(\frac{c-y}{y-l}\bigg)^{\omega}\bigg)^{-1} \cdot \mathds{1}_{[l,c]} (y) + \bigg(1+\bigg(\frac{r-y}{y-c}\bigg)^{-\omega}\bigg)^{-1} \cdot \mathds{1}_{(c,r]} (y)
\end{equation} 

\noindent where $-\infty < l < c < r < +\infty$ and $\omega \in \mathbb{R}^{+}_{0}$. The fuzziness of the set is controlled by the parameter $\omega$, which provides an intensification ($\omega<1$) or a reduction ($\omega>1$) of its shape. {The versatility of such a fuzzy number offers a way to integrate different sources of uncertainty into a unique formal representation. Note that the parameter $\omega$ allows for increasing or decreasing the overall fuzziness of the set \cite{toth2019applying}. Moreover, four-parameter fuzzy numbers require just one intensification parameter (e.g., differently from \cite{nasibov2008nearest}), providing a balance between flexibility and complexity}. {According to the inverted-U effect between response times and  responses on Likert scales, raters tend to show longer response times especially with middle-scale responses \cite{casey2001validating,mignault2008inverted}. In this context,} $\omega$ can be modulated so that longer response times---which are usually provided by raters who are very hesitant about their final choices---produce an intensification of the fuzziness whereas shorter response times --that are usually provided by raters who are quite sure about their final choices--produce a reduction of the fuzziness.  As a result, the fuzziness of the set can be interpreted as a proxy for the rater's decision uncertainty. Figure \ref{fig1} shows an example of the relationship between fuzziness and the $\omega$ parameter. 

\subsection{From IRTree to fIRTree}\label{sec2_2}

\begin{figure}[H]
	\centering
	{\resizebox{4.5cm}{!}{\begin{tikzpicture}[auto,vertex1/.style={draw,circle},vertex2/.style={draw,rectangle}]
			\node[vertex1,minimum size=1cm] (eta1) {$Z_1$};
			\node[vertex1,minimum size=1cm,below right= 1.5cm of eta1] (eta2) {$Z_2$};
			\node[vertex2,below left= 1.5cm of eta1] (y1) {$Y=1$};
			\node[vertex2,below left= 1.5cm of eta2] (y0) {$Y=0$};
			\node[vertex2,below right= 1.5cm of eta2] (y2) {$Y=2$};	
			\draw[-latex] (eta1) -- node[right=0.5 of eta1] {} (eta2);
			\draw[-latex] (eta1) -- node[left=0.5 of eta1] {} (y1);
			\draw[-latex] (eta2) -- node[right=0.5 of eta2] {} (y2);
			\draw[-latex] (eta2) -- node[left=0.5 of eta2] {} (y0);			
\end{tikzpicture}}}%
	\qquad
	{\resizebox{5.2cm}{!}{\begin{tikzpicture}[auto,vertex1/.style={draw,circle},vertex2/.style={draw,rectangle}]
		
			\node[vertex1,minimum size=1cm] (eta1) {$A$};
			\node[vertex1,minimum size=1cm,below right=2cm of eta1] (eta3) {$H$};
			\node[vertex1,minimum size=1cm,below left=2cm of eta1] (eta2) {$L$};
			\node[vertex2,below left=0.5cm of eta2] (Y1) {$Y=1$}; \node[vertex2,below right=0.5cm of eta2] (Y2) {$Y=2$};
			\node[vertex2,below left=0.5cm of eta3] (Y3) {$Y=3$}; \node[vertex2,below right=0.5cm of eta3] (Y4) {$Y=4$};

			\draw[-latex] (eta1) -- node[right=0.5 of eta1] {} (eta3);
			\draw[-latex] (eta1) -- node[right=0.5 of eta1] {} (eta2);
			\draw[-latex] (eta2) -- node[right=0.5 of eta1] {} (Y1); 
			\draw[-latex] (eta2) -- node[right=0.5 of eta1] {} (Y2);
			\draw[-latex] (eta3) -- node[right=0.5 of eta1] {} (Y3); 
			\draw[-latex] (eta3) -- node[right=0.5 of eta1] {} (Y4);
			\end{tikzpicture}}}
	\caption{Example of IRTree of a rating scale with three (e.g., 0: Disagree; 1: Neutral;  2: Agree) or four (e.g., 1: Strongly Disagree; 2: Disagree; 3: Agree; 4: Strongly Agree) response categories.\label{fig2}} 
\end{figure}
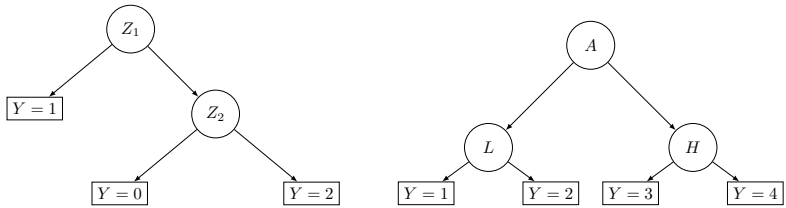

Item Response Theory (IRT) trees are conditional linear models that represent ratings data in terms of binary trees. In general, the tree formalizes the rating response process as a sequence of intermediate nodes, each of which corresponds to a cognitive decision, and end-nodes, which codify the possible final choices or answers. Figure~\ref{fig2} depicts two examples of IRTree models for a rating scale with three and four response categories. In particular, the first tree (Figure \ref{fig2}a) formalizes a typical situation where the rater first decides whether he/she is neutral about the item/question being assessed ($Z_1$) and then he/she decides about the strength of the agreement or disagreement ($Z_2$). As a byproduct of the binary structure, the probability of a final response (e.g., $Y=1$: Neutral) can be computed by multiplying the probabilities of each branch (e.g., $P(Y=1)=P(Z_1=0;\boldsymbol{\theta})$).

By adopting the IRT parametrization, the parameter array $\boldsymbol{\theta}$ can be defined as to contain rater-specific and item-specific parameters. Thus, the probability to agree or disagree with an item/question can be represented as a function of a rater's latent trait and the specific content of the item \cite{Boeck_2012}. More formally, let $i \in \{1,\ldots,I\}$ and $j \in \{1,\ldots,J\}$ be the indices for raters and items, respectively. Then, the final response variable $Y_{ij} \in \{1,\ldots,m,\ldots,M\} \subset\mathbb N$ ($M$ is the maximum number of response categories) can be written as a function of $N$ binary variables $Z_{ijn} \in \{0,1\}$, with $n \in \{1,\ldots,N\}$ denoting the nodes of the tree. For instance, in Figure~\ref{fig2}a, the final response $Y_{ij}=0$ corresponds to $Z_{ij2} = 0$. For a generic pair $(i,j)$ of data, the IRTree model consists of the following equations:
\begin{align}
	& \boldsymbol\eta_{i} \sim \mathcall N_N(\mathbf 0,\boldsymbol{\Sigma}_{\eta}),\label{eq1a}\\
	& \pi_{ijn} = \Prob{Z_{ijn} = 1; \boldsymbol{\theta}_n} = \frac{\exp\left(\eta_{in}+\alpha_{jn}\right)}{1+\exp\left(\eta_{in}+\alpha_{jn}\right)},\label{eq1c}\\
	& Z_{ijn} \sim \mathcall Ber(\pi_{ijn}),\label{eq1d}
\end{align}

\noindent where $\boldsymbol{\theta}_n = \{\boldsymbol{\alpha}_j, \boldsymbol{\beta}_i\}$, with the arrays $\boldsymbol{\alpha}_j \in \mathbb R^N$ and $\boldsymbol{\eta}_i \in \mathbb R^N$ denoting the easiness of the item and the rater's latent trait. As is usual in IRT models, latent traits for each node are modeled using a $N$-variate centered Gaussian distribution with covariance matrix $\boldsymbol{\Sigma}_\eta$. In this representation, the probability for a generic rating response can be computed as:
\begin{align}\label{eq2}
	\Prob{Y_{ij}=m} & = \prod_{n=1}^N \Prob{Z_{ijn} = t_{mn};\boldsymbol{\theta}_n}^{t_{mn}} \nonumber\\
	& = \prod_{n=1}^N \left(\frac{\exp\left(\eta_{in}+\alpha_{jn}\right)t_{mn}}{1+\exp\left(\eta_{in}+\alpha_{jn}\right)}\right)^{\delta_{mn}}
\end{align}

\noindent where $t_{mn}\in\{0,1,\text{NA}\}$ is the entry of the mapping matrix $T_{M\times N}$ with $t_{mn}=1$ indicating a connection from the $m$-th response category to the $n$-th node, $t_{mn}=0$ and $t_{mn}=\text{NA}$ indicating no connection, whereas $\delta_{mn}=0$ if $t_{mn}=\text{NA}$ and $\delta_{mn}=1$ otherwise. 

The parameters $\boldsymbol{\alpha}_1,\ldots,\boldsymbol{\alpha}_J$ and $\boldsymbol{\Sigma}_{\eta}$ of IRTree models can be estimated either by means of standard methods used for generalized linear mixed models---for instance, restricted or marginal maximum likelihood \cite{de2011estimation,Boeck_2012}---or via expectation maximization-based algorithms~\cite{Jeon_2015}.

The fIRTree procedure relies on the use of the estimated array of parameters $\boldsymbol{\hat\theta}$ and the estimated transition probabilities $\ProbEst{Z_{ij1}},\ldots,\ProbEst{Z_{ijn}},\ldots,\ProbEst{Z_{ijN}}$. Further technical details about the connection between IRTree and fIRTree can be found in \cite{calcagni2021firtree,calcagni2022psychometric}. More generally, the input of entire procedure consists of the $I\times J$ matrices of crisp rating responses $\mathbf Y$ and responses times $\mathbf T$ whereas the output is an array of fuzzy data $\mathbf{\widetilde{Y}}$. Note that, the entry $\tilde{y}_{ij}$ of $\mathbf{\widetilde{Y}}$ is the 4-tuple $\{c_{ij},l_{ij},r_{ij}, \omega_{ij}\}$. More in details, for each pair $(i,j)$, fIRTree requires the following steps:

\begin{enumerate}
	\item Define and fit an IRTtree model to $\boldsymbol{Y}_{I\times J}$ in order to get $\boldsymbol{\hat \eta}_{N\times 1}$ and $\boldsymbol{\hat \alpha}_{N\times 1}$
	
	\item Plug-in $\boldsymbol{\hat \eta}_{N\times 1}$ and $\boldsymbol{\hat \alpha}_{N\times 1}$ into Eq.~\eqref{eq2} to get the estimated probability distribution $\ProbEst{Y=1},\ldots,\ProbEst{Y=m},\ldots,\ProbEst{Y=M}$
	
	\item Compute mode $c_{ij}$ and precision $s_{ij}$ of the fuzzy number $\tilde y_{ij}$ via the following equalities:
	\begin{align}
		& c_{ij} = \sum_{y=1}^{M} y\cdot\ProbEst{Y=y}\\
		& s_{ij}  = \sum_{y=1}^{M} (y-c_{ij})^2\cdot \ProbEst{Y=y}
	\end{align}
	
	\item Compute left and right bounds using link equations:
	\begin{align}
		& {l_{ij}}=c_{ij}-h_2\\
		& {r_{ij}}=c_{ij}-h_2+h_1\\
		& \text{ where: }~ h_1 = \sqrt{3.5v_{ij}-3(c_{ij}-\mu_{ij})^2} \\
		& \quad\quad\quad\quad h_2 = \frac{1}{2}(h_1+3c_{ij}-3\mu_{ij})
	\end{align}
	
	\item Compute the fuzzy membership function:
	\begin{equation}
		\mu_{ij} = (1+c_{ij}s_{ij})\big/(2+s_{ij})
	\end{equation}
	
	\item Compute the intensification parameter:
	
	\begin{equation}\label{eq5}
		\omega_{ij}= \hat{F}(Mdn(\mathbf{t}_{\cdot j})) - \hat{F}(t_{ij}) + 1
	\end{equation}
	
\end{enumerate} 

\noindent {Note} that in Equation~\eqref{eq5} the intensification parameter $\omega_{ij}$ is computed using the response times $t_{ij}$ of a subject $i$ to an item $j$, in order to model the uncertainty of the response process. As suggested by \cite{calcagni2014dynamic}, $\hat{F}$ is the empirical cumulative distribution function of the observed response times $\mathbf{t}_{\cdot j}=(t_{1j}, t_{2j}, \dots , t_{ij})$ for the $j$-th item/question whereas $Mdn(\mathbf{t}_{\cdot j})$ is the median of the vector $\mathbf{t}_{\cdot j}$.

Note that Equation~\eqref{eq5} can be interpreted in light of the findings provided by \cite{mignault2008inverted}. In particular, when the $i$-th rater shows a response time such that $t_{ij}<Mdn(\mathbf{t}_{\cdot j})$ then the uncertainty affecting the response process decreases as a function of the fuzziness of the set ($\omega_{ij}>1$). Conversely, the opposite case occur when $t_{ij}>Mdn(\mathbf{t}_{\cdot j})$.

\subsection{Fuzzy Normal linear model with crisp predictors}\label{sec2_3}

Let $\boldsymbol{Y} = \{Y_1,\ldots, Y_i,\ldots, Y_n\}$ be a random sample and $\{X_1,\ldots,X_j,\ldots,X_J\}$ a set of non random variables (i.e., covariates). The $i$-th observation of the random sample is associated with a specific set of covariates $\mathbf x_{i} = \{\mathbf x_{i1},\ldots,\mathbf x_{ij},\ldots,\mathbf x_{iJ}\}$ so that the sample consist of paired observations $\{(Y_1,x_{1j}),\ldots,(Y_i,x_{ij}),\ldots,(Y_n,x_{nj})\}$ for $j=1,\ldots,J$. In order to evaluate whether the outcomes $Y_i$ are linearly related to the covariates $\mathbf x_i$, a normal linear model can be used:
\begin{align}\label{eq6}
	& Y_i \sim \mathcall N(\mu_i,\sigma^2_i) \\\nonumber
	& \mu_i = \beta_0 + \mathbf{x}_i\boldsymbol{\beta}\\
	& \sigma^2_i = \sigma^2\nonumber
\end{align}
where $\{\beta_0,\boldsymbol{\beta}\}\in\mathbb R\times \mathbb R^{J}$ and $\sigma^2\in\mathbb R^+$ is constant over observations (homoscedasticity).  The Likelihood function $\mathcall L(\boldsymbol{\theta};\mathbf{y})$ for the Normal model in Equation~\eqref{eq6} is: 
\begin{equation}
	\mathcall L(\boldsymbol{\theta};\mathbf{y}) =  -\frac{n}{2}\ln(2\pi)-\frac{n}{2}\ln(\sigma^2)-\frac{1}{2\sigma^2} \left(\mathbf{y}-\mathbf{X}\boldsymbol{\beta}\right)^T \left(\mathbf{y}-\mathbf{X}\boldsymbol{\beta}\right)
\end{equation}

\noindent {The} array of parameters is $\boldsymbol{\theta}=\{\beta_0,\boldsymbol{\beta},\sigma^2\}$. In the context of fIRTree data, the random outcomes are formalized in terms of fuzzy observations and a sample of fuzzy data $\mathbf{\tilde y} = \{\tilde y_1,\ldots,\tilde y_i,\ldots,\tilde y_n\}$ is available instead of $\mathbf y$. In this context, the researcher is dealing with two sources of uncertainty: (i) the random variation due to the sampling process, which is codified by $\sigma^2$; (ii) the non-random subjective uncertainty due to the response process, which is codified by the fuzzy datum $\tilde y_i$. As the goal of the statistical modeling still remains the inference of the linear relationship between the outcome $Y_i$ and the predictors $\mathbf x_i$, we need to filter-out the fuzzy component from the data. To this end, the minimum inaccuracy principle can be minimized \cite{corral1984minimun}:
\begin{align*}
	& \mathcall I(\boldsymbol{\theta};\mathbf{\tilde y}) =\sum_{i=1}^{n} ~\int \xi_{\tilde{y}_i}^*(y) \ln \mathcall L(\boldsymbol{\theta};y) ~dy
\end{align*}

\noindent with $\mathcall L(\boldsymbol{\theta};y)$ being the likelihood function for the normal model in Equation~\eqref{eq6}, whereas $\fuzzysetij{y}{i}^*(y)$ is the standardized version of the fuzzy set $\tilde{y}_{i}$, which is in turn obtained by the following calculus:
\begin{equation*}
	\xi_{\tilde y_1,\ldots,\tilde y_n}(y_1,\ldots,y_n) = \frac{\xi_{\tilde y_1,\ldots,\tilde y_n}^*(y_1,\ldots,y_n)}{\idotsint\limits_{}\xi_{\tilde y_1,\ldots,\tilde y_n}(y_1,\ldots,y_n)~dy_1\cdots dy_n}
\end{equation*}

\noindent As usual, the parameters $\boldsymbol{\theta}$ can be obtained by solving the score equation:
\begin{align*}
	\frac{\partial}{\partial \boldsymbol{\theta}}\mathcall I(\boldsymbol{\theta};\mathbf{\tilde y}) = -\sum_{i=1}^{n} ~\int\xi_{\tilde{y}_i}^*(y_i) \frac{\partial}{\partial\boldsymbol{\theta}}\ln \mathcall L(\boldsymbol{\theta};\mathbf{y}) ~dy= \mathbf{0}_{J+2},
\end{align*}
{In} this case, the solutions of the minimization problem can be obtained numerically (e.g., via L-BFGS algorithm).

\section{Application}\label{sec3}

The aim of this section is to provide an application of the fuzzy Normal linear model which has been applied on a real case study involving the Depression, Anxiety and Stress Scale (DASS) \cite{parkitny2010depression}. The data as well as the algorithms used throughout this article are freely available at: \url{https://github.com/niccolocao/Fuzzy-Normal-Model}.

\subsection{Data and variables}\label{sec3_1}

The dataset refers to a sample of $n=160$ participants ($28.7\%$ women with mean age of 22.64 years, std. deviation of 7.43) and $J=14$ items/questions from the Depression Scale of the DASS inventory. The observed data consist of responses on a four-point rating scale along with the response times (in ms). As typical for these studies, answers with response times over and under two standard deviations from the mean have been removed.

The matrices $\mathbf Y_{160\times 14}$ and $\mathbf T_{160\times 14}$ of the crisp rating responses and times have been used as input of the fIRTree procedure (see Section \ref{sec2_2}), which has produced as output the matrices $\mathbf C_{160\times 14}$, $\mathbf L_{160\times 14}$, $\mathbf R_{160\times 14}$, $\mathbf W_{160\times 14}$ of fuzzy parameters. For the sake of simplicity, the linear decision tree with three nodes has been used (see {Figure} \ref{fig3}).  
In this case, the response process is formalized by means of three nodes, namely node $A$ for a first agreement/disagreement toward the item being assessed, node $M$ for a moderate level of agreement/disagreement, and node $E$ for the selection of extreme response categories. The IRTree model has been defined using the R library 
\texttt{IRTrees} \cite{de2012irtrees} whereas model parameters $\boldsymbol\alpha$ and $\boldsymbol{\eta}$ have been estimated using the R library \texttt{glmmTMB} \cite{brooks2017glmm}. Finally, estimated centers, left/right bounds, and intensification parameters of fuzzy numbers have been averaged in order to obtain a composite indicator for depression \cite{yu2007fuzzy}. Figure \ref{fig4} shows the histograms of the composite fuzzy indicator w.r.t. centers, left/right spreads, and intensification parameters. Note that the magnitude of left/right spreads (Figure~\ref{fig4}b,c) shows that a certain level of fuzziness is present in ratings data, whereas the intensification parameter (Figure~\ref{fig3}d) shows a higher concentration of values closed to one ($21.25\%$, reference range: $[0.95,1.05]$). The fuzzy indicator \texttt{depression} has been considered as the response variable of the next statistical models. Three crisp predictors have been used as follows: (a) \texttt{religiousness}, a categorical variable with two categories $\{\texttt{Yes}, \texttt{No}\}$; (ii) \texttt{emotional\_stability}, a compound indicator about personality \cite{gosling2003very}, which has been derived from the Ten Item Personality Inventory~\cite{gosling2003very} as a convex combination of the items and Cronbach's $\alpha$. {{The following} formula has been used for the $\alpha$-based composite indicator: 
\begin{align}
	\texttt{emotional\_stability}_i&=\alpha \times \sum_{j=1}^{J}\texttt{emotional\_stability}_{ij}+ \\
	& + (1-\alpha)\times mean\left(\sum_{j=1}^{J} \texttt{emotional\_stability}_{ij}\right)\nonumber
\end{align}	
	
\noindent {{Note} that in this context the $\alpha$ coefficient is used to create a crisp composite indicator where the contribution of the single item is weighted by the overall internal consistency of the scale they belong to. This should not be confused with the fuzzy $\alpha$ coefficient provided by \cite{lubiano2021fuzzy}, which is instead used to assess the internal consistency of a fuzzy scale}}; (iii) \texttt{university}, a categorical variable with two categories $\{\texttt{Yes}, \texttt{No}\}$, with $\texttt{Yes}$ indicating the case where participants have reached the university education level.

\begin{figure}[H]
	\centering
	\resizebox{4.8cm}{!}{
		\begin{tikzpicture}[auto,vertex1/.style={draw,circle},vertex2/.style={draw,rectangle}]
			\node[vertex1,minimum size=1cm] (eta1) {$A$};
			\node[vertex1,minimum size=1cm,below right=1cm of eta1] (eta2) {$M$};
			\node[vertex2,below left=0.5cm of eta1] (Y0) {$Y=0$};
			\node[vertex1,minimum size=1cm,below right=1cm of eta2] (eta3) {$E$};

			\node[vertex2,below left=0.5cm of eta2] (Y1) {$Y=1$}; 
			\node[vertex2,below right=0.5cm of eta3] (Y3) {$Y=3$};
			\node[vertex2,below left=0.5cm of eta3] (Y2) {$Y=2$};
			
			\draw[-latex] (eta1) -- node[right=0.5 of eta1] {} (eta2);
			\draw[-latex] (eta1) -- node[right=0.5 of eta1] {} (Y0); 
			\draw[-latex] (eta2) -- node[right=0.5 of eta1] {} (eta3);
			\draw[-latex] (eta2) -- node[right=0.5 of eta1] {} (Y1);
			\draw[-latex] (eta3) -- node[right=0.5 of eta1] {} (Y2); 
			\draw[-latex] (eta3) -- node[right=0.5 of eta1] {} (Y3);
\end{tikzpicture}
	}
	\caption{{Application}: Linear decision tree for the four response categories. Note that $A$, $M$, $E$ denote the decision nodes of disagreement, moderate agreement/disagreement, extreme agreement, respectively.\label{fig3}} 
\end{figure}
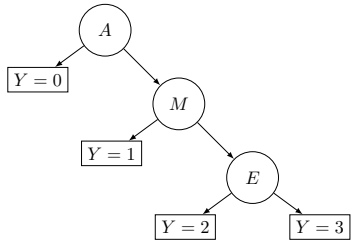

\begin{figure}[H]
	\hspace{-1.65cm}
	{%
		\resizebox{5.75cm}{!}{
\begin{tikzpicture}[x=1pt,y=1pt]
\definecolor{fillColor}{RGB}{255,255,255}
\path[use as bounding box,fill=fillColor,fill opacity=0.00] (0,0) rectangle (614.29,758.83);
\begin{scope}
\path[clip] (  0.00,  0.00) rectangle (614.29,758.83);
\definecolor{drawColor}{RGB}{0,0,0}

\path[draw=drawColor,line width= 0.4pt,line join=round,line cap=round] (156.58, 80.67) -- (457.71, 80.67);

\path[draw=drawColor,line width= 0.4pt,line join=round,line cap=round] (156.58, 80.67) -- (156.58, 74.67);

\path[draw=drawColor,line width= 0.4pt,line join=round,line cap=round] (206.77, 80.67) -- (206.77, 74.67);

\path[draw=drawColor,line width= 0.4pt,line join=round,line cap=round] (256.96, 80.67) -- (256.96, 74.67);

\path[draw=drawColor,line width= 0.4pt,line join=round,line cap=round] (307.15, 80.67) -- (307.15, 74.67);

\path[draw=drawColor,line width= 0.4pt,line join=round,line cap=round] (357.33, 80.67) -- (357.33, 74.67);

\path[draw=drawColor,line width= 0.4pt,line join=round,line cap=round] (407.52, 80.67) -- (407.52, 74.67);

\path[draw=drawColor,line width= 0.4pt,line join=round,line cap=round] (457.71, 80.67) -- (457.71, 74.67);

\node[text=drawColor,anchor=base,inner sep=0pt, outer sep=0pt, scale=  2.75] at (156.58, 50.67) {1.0};

\node[text=drawColor,anchor=base,inner sep=0pt, outer sep=0pt, scale=  2.75] at (256.96, 50.67) {2.0};

\node[text=drawColor,anchor=base,inner sep=0pt, outer sep=0pt, scale=  2.75] at (357.33, 50.67) {3.0};

\node[text=drawColor,anchor=base,inner sep=0pt, outer sep=0pt, scale=  2.75] at (457.71, 50.67) {4.0};

\path[draw=drawColor,line width= 0.4pt,line join=round,line cap=round] (152.94, 96.36) -- (152.94,598.23);

\path[draw=drawColor,line width= 0.4pt,line join=round,line cap=round] (152.94, 96.36) -- (146.94, 96.36);

\path[draw=drawColor,line width= 0.4pt,line join=round,line cap=round] (152.94,263.65) -- (146.94,263.65);

\path[draw=drawColor,line width= 0.4pt,line join=round,line cap=round] (152.94,430.94) -- (146.94,430.94);

\path[draw=drawColor,line width= 0.4pt,line join=round,line cap=round] (152.94,598.23) -- (146.94,598.23);

\node[text=drawColor,rotate= 90.00,anchor=base,inner sep=0pt, outer sep=0pt, scale=  2.75] at (130.14, 96.36) {0};

\node[text=drawColor,rotate= 90.00,anchor=base,inner sep=0pt, outer sep=0pt, scale=  2.75] at (130.14,263.65) {5};

\node[text=drawColor,rotate= 90.00,anchor=base,inner sep=0pt, outer sep=0pt, scale=  2.75] at (130.14,430.94) {10};

\node[text=drawColor,rotate= 90.00,anchor=base,inner sep=0pt, outer sep=0pt, scale=  2.75] at (130.14,598.23) {15};
\end{scope}
\begin{scope}
\path[clip] (144.54, 72.27) rectangle (469.75,722.70);
\definecolor{drawColor}{RGB}{0,0,0}
\definecolor{fillColor}{RGB}{255,255,255}

\path[draw=drawColor,line width= 0.4pt,line join=round,line cap=round,fill=fillColor] (156.58, 96.36) rectangle (176.66,397.48);

\path[draw=drawColor,line width= 0.4pt,line join=round,line cap=round,fill=fillColor] (176.66, 96.36) rectangle (196.73,330.57);

\path[draw=drawColor,line width= 0.4pt,line join=round,line cap=round,fill=fillColor] (196.73, 96.36) rectangle (216.81,430.94);

\path[draw=drawColor,line width= 0.4pt,line join=round,line cap=round,fill=fillColor] (216.81, 96.36) rectangle (236.88,497.86);

\path[draw=drawColor,line width= 0.4pt,line join=round,line cap=round,fill=fillColor] (236.88, 96.36) rectangle (256.96,665.15);

\path[draw=drawColor,line width= 0.4pt,line join=round,line cap=round,fill=fillColor] (256.96, 96.36) rectangle (277.03,464.40);

\path[draw=drawColor,line width= 0.4pt,line join=round,line cap=round,fill=fillColor] (277.03, 96.36) rectangle (297.11,564.78);

\path[draw=drawColor,line width= 0.4pt,line join=round,line cap=round,fill=fillColor] (297.11, 96.36) rectangle (317.18,330.57);

\path[draw=drawColor,line width= 0.4pt,line join=round,line cap=round,fill=fillColor] (317.18, 96.36) rectangle (337.26,464.40);

\path[draw=drawColor,line width= 0.4pt,line join=round,line cap=round,fill=fillColor] (337.26, 96.36) rectangle (357.33,497.86);

\path[draw=drawColor,line width= 0.4pt,line join=round,line cap=round,fill=fillColor] (357.33, 96.36) rectangle (377.41,230.19);

\path[draw=drawColor,line width= 0.4pt,line join=round,line cap=round,fill=fillColor] (377.41, 96.36) rectangle (397.48,297.11);

\path[draw=drawColor,line width= 0.4pt,line join=round,line cap=round,fill=fillColor] (397.48, 96.36) rectangle (417.56,497.86);

\path[draw=drawColor,line width= 0.4pt,line join=round,line cap=round,fill=fillColor] (417.56, 96.36) rectangle (437.63,430.94);

\path[draw=drawColor,line width= 0.4pt,line join=round,line cap=round,fill=fillColor] (437.63, 96.36) rectangle (457.71,698.61);
\end{scope}
\begin{scope}
\path[clip] (  0.00,  0.00) rectangle (614.29,758.83);
\definecolor{drawColor}{RGB}{0,0,0}

\node[text=drawColor,anchor=base,inner sep=0pt, outer sep=0pt, scale=  3.00] at (307.15, 14.67) {Centers};

\node[text=drawColor,rotate= 90.00,anchor=base,inner sep=0pt, outer sep=0pt, scale=  3.00] at ( 94.14,397.48) {Frequency};
\end{scope}
\end{tikzpicture}}%
	}%
	\hspace{-2.15cm}
	{%
		\resizebox{5.75cm}{!}{
\begin{tikzpicture}[x=1pt,y=1pt]
\definecolor{fillColor}{RGB}{255,255,255}
\path[use as bounding box,fill=fillColor,fill opacity=0.00] (0,0) rectangle (614.29,758.83);
\begin{scope}
\path[clip] (  0.00,  0.00) rectangle (614.29,758.83);
\definecolor{drawColor}{RGB}{0,0,0}

\path[draw=drawColor,line width= 0.4pt,line join=round,line cap=round] (156.58, 80.67) -- (457.71, 80.67);

\path[draw=drawColor,line width= 0.4pt,line join=round,line cap=round] (156.58, 80.67) -- (156.58, 74.67);

\path[draw=drawColor,line width= 0.4pt,line join=round,line cap=round] (206.77, 80.67) -- (206.77, 74.67);

\path[draw=drawColor,line width= 0.4pt,line join=round,line cap=round] (256.96, 80.67) -- (256.96, 74.67);

\path[draw=drawColor,line width= 0.4pt,line join=round,line cap=round] (307.15, 80.67) -- (307.15, 74.67);

\path[draw=drawColor,line width= 0.4pt,line join=round,line cap=round] (357.33, 80.67) -- (357.33, 74.67);

\path[draw=drawColor,line width= 0.4pt,line join=round,line cap=round] (407.52, 80.67) -- (407.52, 74.67);

\path[draw=drawColor,line width= 0.4pt,line join=round,line cap=round] (457.71, 80.67) -- (457.71, 74.67);

\node[text=drawColor,anchor=base,inner sep=0pt, outer sep=0pt, scale=  2.75] at (156.58, 50.67) {0.0};

\node[text=drawColor,anchor=base,inner sep=0pt, outer sep=0pt, scale=  2.75] at (256.96, 50.67) {0.4};

\node[text=drawColor,anchor=base,inner sep=0pt, outer sep=0pt, scale=  2.75] at (357.33, 50.67) {0.8};

\node[text=drawColor,anchor=base,inner sep=0pt, outer sep=0pt, scale=  2.75] at (457.71, 50.67) {1.2};

\path[draw=drawColor,line width= 0.4pt,line join=round,line cap=round] (152.94, 96.36) -- (152.94,643.86);

\path[draw=drawColor,line width= 0.4pt,line join=round,line cap=round] (152.94, 96.36) -- (146.94, 96.36);

\path[draw=drawColor,line width= 0.4pt,line join=round,line cap=round] (152.94,187.61) -- (146.94,187.61);

\path[draw=drawColor,line width= 0.4pt,line join=round,line cap=round] (152.94,278.86) -- (146.94,278.86);

\path[draw=drawColor,line width= 0.4pt,line join=round,line cap=round] (152.94,370.11) -- (146.94,370.11);

\path[draw=drawColor,line width= 0.4pt,line join=round,line cap=round] (152.94,461.36) -- (146.94,461.36);

\path[draw=drawColor,line width= 0.4pt,line join=round,line cap=round] (152.94,552.61) -- (146.94,552.61);

\path[draw=drawColor,line width= 0.4pt,line join=round,line cap=round] (152.94,643.86) -- (146.94,643.86);

\node[text=drawColor,rotate= 90.00,anchor=base,inner sep=0pt, outer sep=0pt, scale=  2.75] at (130.14, 96.36) {0};

\node[text=drawColor,rotate= 90.00,anchor=base,inner sep=0pt, outer sep=0pt, scale=  2.75] at (130.14,187.61) {5};

\node[text=drawColor,rotate= 90.00,anchor=base,inner sep=0pt, outer sep=0pt, scale=  2.75] at (130.14,278.86) {10};

\node[text=drawColor,rotate= 90.00,anchor=base,inner sep=0pt, outer sep=0pt, scale=  2.75] at (130.14,370.11) {15};

\node[text=drawColor,rotate= 90.00,anchor=base,inner sep=0pt, outer sep=0pt, scale=  2.75] at (130.14,461.36) {20};

\node[text=drawColor,rotate= 90.00,anchor=base,inner sep=0pt, outer sep=0pt, scale=  2.75] at (130.14,552.61) {25};

\node[text=drawColor,rotate= 90.00,anchor=base,inner sep=0pt, outer sep=0pt, scale=  2.75] at (130.14,643.86) {30};
\end{scope}
\begin{scope}
\path[clip] (144.54, 72.27) rectangle (469.75,722.70);
\definecolor{drawColor}{RGB}{0,0,0}
\definecolor{fillColor}{RGB}{255,255,255}

\path[draw=drawColor,line width= 0.4pt,line join=round,line cap=round,fill=fillColor] (156.58, 96.36) rectangle (181.68,406.61);

\path[draw=drawColor,line width= 0.4pt,line join=round,line cap=round,fill=fillColor] (181.68, 96.36) rectangle (206.77,278.86);

\path[draw=drawColor,line width= 0.4pt,line join=round,line cap=round,fill=fillColor] (206.77, 96.36) rectangle (231.87,315.36);

\path[draw=drawColor,line width= 0.4pt,line join=round,line cap=round,fill=fillColor] (231.87, 96.36) rectangle (256.96,224.11);

\path[draw=drawColor,line width= 0.4pt,line join=round,line cap=round,fill=fillColor] (256.96, 96.36) rectangle (282.05,151.11);

\path[draw=drawColor,line width= 0.4pt,line join=round,line cap=round,fill=fillColor] (282.05, 96.36) rectangle (307.15,424.86);

\path[draw=drawColor,line width= 0.4pt,line join=round,line cap=round,fill=fillColor] (307.15, 96.36) rectangle (332.24,698.61);

\path[draw=drawColor,line width= 0.4pt,line join=round,line cap=round,fill=fillColor] (332.24, 96.36) rectangle (357.33,479.61);

\path[draw=drawColor,line width= 0.4pt,line join=round,line cap=round,fill=fillColor] (357.33, 96.36) rectangle (382.43,516.11);

\path[draw=drawColor,line width= 0.4pt,line join=round,line cap=round,fill=fillColor] (382.43, 96.36) rectangle (407.52,260.61);

\path[draw=drawColor,line width= 0.4pt,line join=round,line cap=round,fill=fillColor] (407.52, 96.36) rectangle (432.62,169.36);

\path[draw=drawColor,line width= 0.4pt,line join=round,line cap=round,fill=fillColor] (432.62, 96.36) rectangle (457.71,151.11);
\end{scope}
\begin{scope}
\path[clip] (  0.00,  0.00) rectangle (614.29,758.83);
\definecolor{drawColor}{RGB}{0,0,0}

\node[text=drawColor,anchor=base,inner sep=0pt, outer sep=0pt, scale=  3.00] at (307.15, 14.67) {Left spreads};

\node[text=drawColor,rotate= 90.00,anchor=base,inner sep=0pt, outer sep=0pt, scale=  3.00] at ( 94.14,397.48) {Frequency};
\end{scope}
\end{tikzpicture}}%
	}%
	\hspace{-2.15cm}
	{%
		\resizebox{5.75cm}{!}{
\begin{tikzpicture}[x=1pt,y=1pt]
\definecolor{fillColor}{RGB}{255,255,255}
\path[use as bounding box,fill=fillColor,fill opacity=0.00] (0,0) rectangle (614.29,758.83);
\begin{scope}
\path[clip] (  0.00,  0.00) rectangle (614.29,758.83);
\definecolor{drawColor}{RGB}{0,0,0}

\path[draw=drawColor,line width= 0.4pt,line join=round,line cap=round] (156.58, 80.67) -- (430.33, 80.67);

\path[draw=drawColor,line width= 0.4pt,line join=round,line cap=round] (156.58, 80.67) -- (156.58, 74.67);

\path[draw=drawColor,line width= 0.4pt,line join=round,line cap=round] (211.33, 80.67) -- (211.33, 74.67);

\path[draw=drawColor,line width= 0.4pt,line join=round,line cap=round] (266.08, 80.67) -- (266.08, 74.67);

\path[draw=drawColor,line width= 0.4pt,line join=round,line cap=round] (320.83, 80.67) -- (320.83, 74.67);

\path[draw=drawColor,line width= 0.4pt,line join=round,line cap=round] (375.58, 80.67) -- (375.58, 74.67);

\path[draw=drawColor,line width= 0.4pt,line join=round,line cap=round] (430.33, 80.67) -- (430.33, 74.67);

\node[text=drawColor,anchor=base,inner sep=0pt, outer sep=0pt, scale=  2.75] at (156.58, 50.67) {0.0};

\node[text=drawColor,anchor=base,inner sep=0pt, outer sep=0pt, scale=  2.75] at (266.08, 50.67) {0.4};

\node[text=drawColor,anchor=base,inner sep=0pt, outer sep=0pt, scale=  2.75] at (375.58, 50.67) {0.8};

\path[draw=drawColor,line width= 0.4pt,line join=round,line cap=round] (152.94, 96.36) -- (152.94,698.61);

\path[draw=drawColor,line width= 0.4pt,line join=round,line cap=round] (152.94, 96.36) -- (146.94, 96.36);

\path[draw=drawColor,line width= 0.4pt,line join=round,line cap=round] (152.94,196.74) -- (146.94,196.74);

\path[draw=drawColor,line width= 0.4pt,line join=round,line cap=round] (152.94,297.11) -- (146.94,297.11);

\path[draw=drawColor,line width= 0.4pt,line join=round,line cap=round] (152.94,397.48) -- (146.94,397.48);

\path[draw=drawColor,line width= 0.4pt,line join=round,line cap=round] (152.94,497.86) -- (146.94,497.86);

\path[draw=drawColor,line width= 0.4pt,line join=round,line cap=round] (152.94,598.23) -- (146.94,598.23);

\path[draw=drawColor,line width= 0.4pt,line join=round,line cap=round] (152.94,698.61) -- (146.94,698.61);

\node[text=drawColor,rotate= 90.00,anchor=base,inner sep=0pt, outer sep=0pt, scale=  2.75] at (130.14, 96.36) {0};

\node[text=drawColor,rotate= 90.00,anchor=base,inner sep=0pt, outer sep=0pt, scale=  2.75] at (130.14,196.74) {5};

\node[text=drawColor,rotate= 90.00,anchor=base,inner sep=0pt, outer sep=0pt, scale=  2.75] at (130.14,297.11) {10};

\node[text=drawColor,rotate= 90.00,anchor=base,inner sep=0pt, outer sep=0pt, scale=  2.75] at (130.14,397.48) {15};

\node[text=drawColor,rotate= 90.00,anchor=base,inner sep=0pt, outer sep=0pt, scale=  2.75] at (130.14,497.86) {20};

\node[text=drawColor,rotate= 90.00,anchor=base,inner sep=0pt, outer sep=0pt, scale=  2.75] at (130.14,598.23) {25};

\node[text=drawColor,rotate= 90.00,anchor=base,inner sep=0pt, outer sep=0pt, scale=  2.75] at (130.14,698.61) {30};
\end{scope}
\begin{scope}
\path[clip] (144.54, 72.27) rectangle (469.75,722.70);
\definecolor{drawColor}{RGB}{0,0,0}
\definecolor{fillColor}{RGB}{255,255,255}

\path[draw=drawColor,line width= 0.4pt,line join=round,line cap=round,fill=fillColor] (156.58, 96.36) rectangle (183.96,256.96);

\path[draw=drawColor,line width= 0.4pt,line join=round,line cap=round,fill=fillColor] (183.96, 96.36) rectangle (211.33,457.71);

\path[draw=drawColor,line width= 0.4pt,line join=round,line cap=round,fill=fillColor] (211.33, 96.36) rectangle (238.71,196.74);

\path[draw=drawColor,line width= 0.4pt,line join=round,line cap=round,fill=fillColor] (238.71, 96.36) rectangle (266.08,357.33);

\path[draw=drawColor,line width= 0.4pt,line join=round,line cap=round,fill=fillColor] (266.08, 96.36) rectangle (293.46,437.63);

\path[draw=drawColor,line width= 0.4pt,line join=round,line cap=round,fill=fillColor] (293.46, 96.36) rectangle (320.83,558.08);

\path[draw=drawColor,line width= 0.4pt,line join=round,line cap=round,fill=fillColor] (320.83, 96.36) rectangle (348.21,638.38);

\path[draw=drawColor,line width= 0.4pt,line join=round,line cap=round,fill=fillColor] (348.21, 96.36) rectangle (375.58,698.61);

\path[draw=drawColor,line width= 0.4pt,line join=round,line cap=round,fill=fillColor] (375.58, 96.36) rectangle (402.96,337.26);

\path[draw=drawColor,line width= 0.4pt,line join=round,line cap=round,fill=fillColor] (402.96, 96.36) rectangle (430.33,216.81);

\path[draw=drawColor,line width= 0.4pt,line join=round,line cap=round,fill=fillColor] (430.33, 96.36) rectangle (457.71,116.44);
\end{scope}
\begin{scope}
\path[clip] (  0.00,  0.00) rectangle (614.29,758.83);
\definecolor{drawColor}{RGB}{0,0,0}

\node[text=drawColor,anchor=base,inner sep=0pt, outer sep=0pt, scale=  3.00] at (307.15, 14.67) {Right spreads};

\node[text=drawColor,rotate= 90.00,anchor=base,inner sep=0pt, outer sep=0pt, scale=  3.00] at ( 94.14,397.48) {Frequency};
\end{scope}
\end{tikzpicture}}%
	}%
	\hspace{-2.15cm}
	{%
		\resizebox{5.75cm}{!}{
\begin{tikzpicture}[x=1pt,y=1pt]
\definecolor{fillColor}{RGB}{255,255,255}
\path[use as bounding box,fill=fillColor,fill opacity=0.00] (0,0) rectangle (614.29,758.83);
\begin{scope}
\path[clip] (  0.00,  0.00) rectangle (614.29,758.83);
\definecolor{drawColor}{RGB}{0,0,0}

\path[draw=drawColor,line width= 0.4pt,line join=round,line cap=round] (186.70, 80.67) -- (427.60, 80.67);

\path[draw=drawColor,line width= 0.4pt,line join=round,line cap=round] (186.70, 80.67) -- (186.70, 74.67);

\path[draw=drawColor,line width= 0.4pt,line join=round,line cap=round] (246.92, 80.67) -- (246.92, 74.67);

\path[draw=drawColor,line width= 0.4pt,line join=round,line cap=round] (307.15, 80.67) -- (307.15, 74.67);

\path[draw=drawColor,line width= 0.4pt,line join=round,line cap=round] (367.37, 80.67) -- (367.37, 74.67);

\path[draw=drawColor,line width= 0.4pt,line join=round,line cap=round] (427.60, 80.67) -- (427.60, 74.67);

\node[text=drawColor,anchor=base,inner sep=0pt, outer sep=0pt, scale=  2.75] at (186.70, 50.67) {0.6};

\node[text=drawColor,anchor=base,inner sep=0pt, outer sep=0pt, scale=  2.75] at (246.92, 50.67) {0.8};

\node[text=drawColor,anchor=base,inner sep=0pt, outer sep=0pt, scale=  2.75] at (307.15, 50.67) {1.0};

\node[text=drawColor,anchor=base,inner sep=0pt, outer sep=0pt, scale=  2.75] at (367.37, 50.67) {1.2};

\node[text=drawColor,anchor=base,inner sep=0pt, outer sep=0pt, scale=  2.75] at (427.60, 50.67) {1.4};

\path[draw=drawColor,line width= 0.4pt,line join=round,line cap=round] (152.94, 96.36) -- (152.94,643.86);

\path[draw=drawColor,line width= 0.4pt,line join=round,line cap=round] (152.94, 96.36) -- (146.94, 96.36);

\path[draw=drawColor,line width= 0.4pt,line join=round,line cap=round] (152.94,187.61) -- (146.94,187.61);

\path[draw=drawColor,line width= 0.4pt,line join=round,line cap=round] (152.94,278.86) -- (146.94,278.86);

\path[draw=drawColor,line width= 0.4pt,line join=round,line cap=round] (152.94,370.11) -- (146.94,370.11);

\path[draw=drawColor,line width= 0.4pt,line join=round,line cap=round] (152.94,461.36) -- (146.94,461.36);

\path[draw=drawColor,line width= 0.4pt,line join=round,line cap=round] (152.94,552.61) -- (146.94,552.61);

\path[draw=drawColor,line width= 0.4pt,line join=round,line cap=round] (152.94,643.86) -- (146.94,643.86);

\node[text=drawColor,rotate= 90.00,anchor=base,inner sep=0pt, outer sep=0pt, scale=  2.75] at (130.14, 96.36) {0};

\node[text=drawColor,rotate= 90.00,anchor=base,inner sep=0pt, outer sep=0pt, scale=  2.75] at (130.14,187.61) {5};

\node[text=drawColor,rotate= 90.00,anchor=base,inner sep=0pt, outer sep=0pt, scale=  2.75] at (130.14,278.86) {10};

\node[text=drawColor,rotate= 90.00,anchor=base,inner sep=0pt, outer sep=0pt, scale=  2.75] at (130.14,370.11) {15};

\node[text=drawColor,rotate= 90.00,anchor=base,inner sep=0pt, outer sep=0pt, scale=  2.75] at (130.14,461.36) {20};

\node[text=drawColor,rotate= 90.00,anchor=base,inner sep=0pt, outer sep=0pt, scale=  2.75] at (130.14,552.61) {25};

\node[text=drawColor,rotate= 90.00,anchor=base,inner sep=0pt, outer sep=0pt, scale=  2.75] at (130.14,643.86) {30};
\end{scope}
\begin{scope}
\path[clip] (144.54, 72.27) rectangle (469.75,722.70);
\definecolor{drawColor}{RGB}{0,0,0}
\definecolor{fillColor}{RGB}{255,255,255}

\path[draw=drawColor,line width= 0.4pt,line join=round,line cap=round,fill=fillColor] (156.58, 96.36) rectangle (186.70,132.86);

\path[draw=drawColor,line width= 0.4pt,line join=round,line cap=round,fill=fillColor] (186.70, 96.36) rectangle (216.81,205.86);

\path[draw=drawColor,line width= 0.4pt,line join=round,line cap=round,fill=fillColor] (216.81, 96.36) rectangle (246.92,351.86);

\path[draw=drawColor,line width= 0.4pt,line join=round,line cap=round,fill=fillColor] (246.92, 96.36) rectangle (277.03,662.11);

\path[draw=drawColor,line width= 0.4pt,line join=round,line cap=round,fill=fillColor] (277.03, 96.36) rectangle (307.15,698.61);

\path[draw=drawColor,line width= 0.4pt,line join=round,line cap=round,fill=fillColor] (307.15, 96.36) rectangle (337.26,552.61);

\path[draw=drawColor,line width= 0.4pt,line join=round,line cap=round,fill=fillColor] (337.26, 96.36) rectangle (367.37,497.86);

\path[draw=drawColor,line width= 0.4pt,line join=round,line cap=round,fill=fillColor] (367.37, 96.36) rectangle (397.48,424.86);

\path[draw=drawColor,line width= 0.4pt,line join=round,line cap=round,fill=fillColor] (397.48, 96.36) rectangle (427.60,242.36);

\path[draw=drawColor,line width= 0.4pt,line join=round,line cap=round,fill=fillColor] (427.60, 96.36) rectangle (457.71,114.61);
\end{scope}
\begin{scope}
\path[clip] (  0.00,  0.00) rectangle (614.29,758.83);
\definecolor{drawColor}{RGB}{0,0,0}

\node[text=drawColor,anchor=base,inner sep=0pt, outer sep=0pt, scale=  3.00] at (307.15, 14.67) {Intensification parameters};

\node[text=drawColor,rotate= 90.00,anchor=base,inner sep=0pt, outer sep=0pt, scale=  3.00] at ( 94.14,397.48) {Frequency};
\end{scope}
\end{tikzpicture}}%
	}%
	\caption{{Application}: Histograms of the parameters of composite fuzzy indicator \texttt{depression}. \textbf{(a)}: Histogram of the centers $c_i$ distribution; \textbf{(b)}: Histogram of left spreads ($l_i-c_i$) distribution; \textbf{(c)}: Histogram of right spreads ($c_i-r_i$) distribution; \textbf{(d)}: Histogram of intensification parameters ($\omega_i$) distribution. Note that the intensification parameter shows a concentration closed to one.\label{fig4}}
\end{figure}
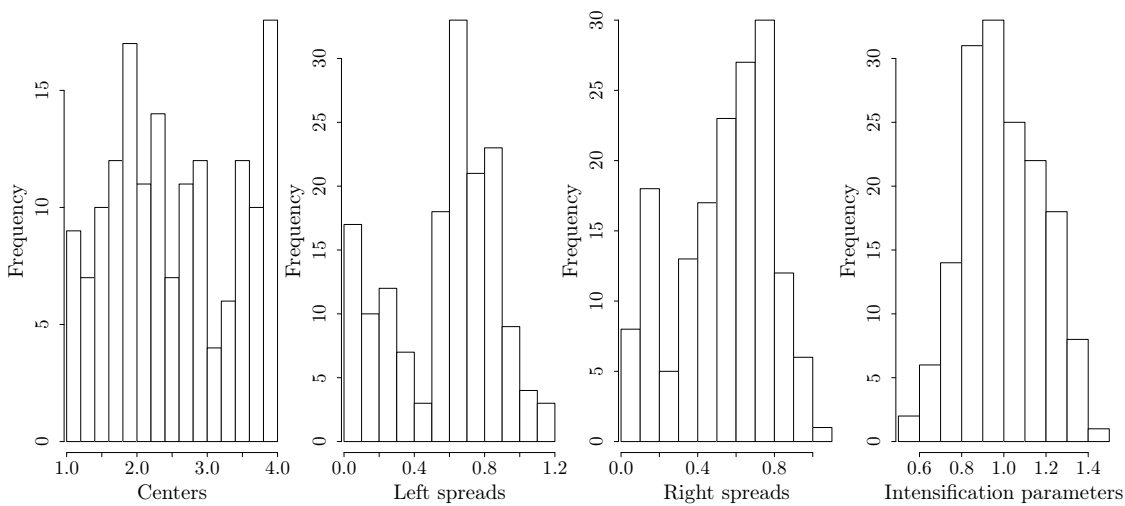

\subsection{Data analysis and results}\label{sec3_2}

In order to compare the results of the fuzzy approach to data analysis as opposed to the standard way of analyzing this data, a Normal linear model and a Log-Normal linear model have been additionally defined and fit on the crisp indicators for depression and response times. In this respect, a traditional data analysis procedure would require to estimate two separate models, one for crisp ratings and a second one for the response times. Table~\ref{tab2} shows the estimated coefficients, standard errors and the associated confidence intervals (CIs) for the two linear models. In particular, the results indicate that \texttt{depression} decreased as a function of \texttt{emotional$\_$stability} $(\hat{\beta}_1=-0.232, \sigma_{\hat{\beta}_1} =0.029, \text{CI}=[-0.289, -0.174] )$, \texttt{religiousness} $(\hat{\beta}_1=-0.146,  \sigma_{\hat{\beta}_1} =0.135, \text{CI}=[-0.412, 0.120]$), with a small decrement from \texttt{No} to \texttt{Yes}, and \texttt{university} $(\hat{\beta}_1=-0.280, \sigma_{\hat{\beta}_1} =0.119, \text{CI}=[-0.516, -0.044]$), with a significant decrease \texttt{No} to \texttt{Yes}. On the other hand, the results of Log-Normal linear model reveal a positive relationship between response times and \texttt{emotional$\_$stability} $(\exp(\hat{\beta}_1)=1.003,\exp(\sigma_{\hat{\beta}_1}) = 1.019, \text{CI}=[-0.034,0.041]$), \texttt{religiousness} $(\exp(\hat{\beta}_1)=1.292,\exp(\sigma_{\hat{\beta}_1}) =1.092, \text{CI}=[0.083,0.429]$), and \texttt{university} $(\exp(\hat{\beta}_1)= 1.176,\exp(\sigma_{\hat{\beta}_1}) =1.081, \text{CI}=[0.008,0.316]$). Considering that participants dedicating a larger amount of time to respond might show a more uncertain response process \cite{casey2001validating}, religious and highly educated participants seem to provide more uncertain responses as opposed to not religious and less educated participants. Overall, the results of traditional data analysis are consistent with the literature (e.g., see \cite{bjelland2008does,koenig2009research,kotov2010linking}). However, as a result of the two-stage data analysis, two issues might affect the overall validity and generalization of the results, namely the need of a correction for multiple testing and the bias that could potentially affect the inferential results of both the linear models. Note that these two limitations derive from the fact that the linear models being used to analyze ratings and response times entail independent inferential tests.

\begin{figure}[h!]
	\hspace{-1cm}
	\centering
	\hspace{-1.5cm}
	\resizebox{10cm}{!}{\input{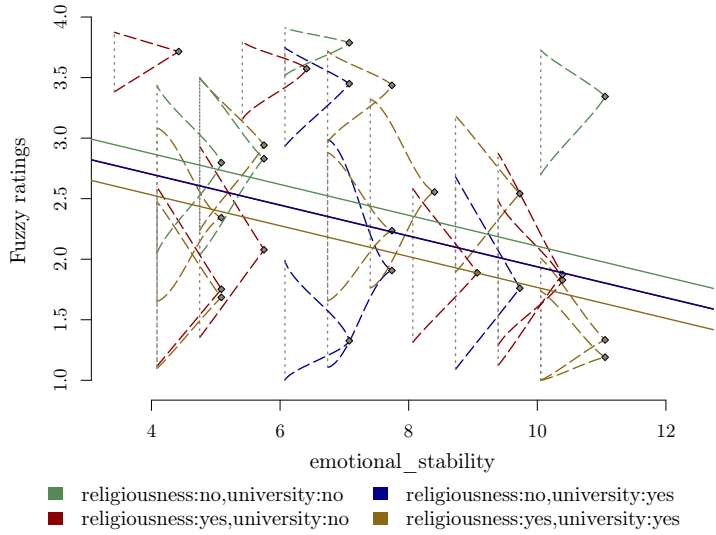}}
	\caption{{Application}: Fitted regression lines against the observed fuzzy ratings as a function of both categorical and continuous predictors. Note that the four regression lines correspond to the four categorical levels of \texttt{religiousness} and \texttt{university} (the line associated with \texttt{religiousness:yes} $\land$ \texttt{university:no} is overlapped with the line of \texttt{religiousness:no} $\land$ \texttt{university:yes}). \label{fig5}}
\end{figure}

On the contrary, fuzzy data analysis provides a unique statistical representation for the analysis and interpretation of both ratings data and response times. Unlike for the previous case, a single (fuzzy) linear Normal model is instead used, where response times have been integrated via the intensification parameter of the fuzzy sets. Table~\ref{tab2} shows the results for this model. Overall, there is a significant negative relationship between \texttt{depression} and \texttt{emotional$\_$stability} $(\hat{\beta}_1=-0.128, \sigma_{\hat{\beta}_1} =0.034, \text{CI}= [-0.195,-0.060])$, with a larger effect of \texttt{religiousness} $(\hat{\beta}_1=-0.170,  \sigma_{\hat{\beta}_1} =0.151, \text{CI}=[-0.468,0.127])$ and \texttt{university} $(\hat{\beta}_1=-0.172, \sigma_{\hat{\beta}_1} =0.133, \text{CI}=[-0.434,0.090])$. Interestingly, the goodness-of-fit indices (i.e., the pseudo-$R^2$s \cite{veall1994evaluating}) of the fuzzy Normal linear model show quite different results from those of the of Normal linear model. Indeed, the pseudo-$R^2$ for the fuzzy Normal linear model (pseudo-$R^2=0.304$) seems to average the indices of the Normal (pseudo-$R^2=0.507$) and Log-Normal (pseudo-$R^2=0.199$) linear models. In order to verify whether the temporal component of the ratings data plays a role in this case, an additional fuzzy Normal linear model has been defined and fit, with the matrix $\mathbf W$ being set equal to one (i.e., $\mathbf W = \mathbf 1_{160\times 14}$). The goodness-of-fit index for this model shows a goodness-of-fit index quite closed to that of the previously estimated model (pseudo-$R^2_{\mathbf W = \mathbf 1}=0.306$, pseudo-$R^2_{\mathbf W \neq \mathbf 1}=0.304$), which would indicate that response times provide a marginal contribution in the analysis of \texttt{depression}. Finally, Figure~\ref{fig5} plots the fitted lines against the observed fuzzy ratings as a function of the predictors.

\begin{table}[!h]
	\small
	\begin{tabular}{lll}
		\toprule
		\textbf{Models}&\quad $\hat{\boldsymbol{\beta}}$ ($\boldsymbol{\sigma}_{\hat{\boldsymbol{\beta}}}$) & (1-$\alpha$\%) \textbf{CI}  \\
		\midrule
		\textbf{Normal Linear Model}: &  &  \\
		Residuals quantiles: Q1: -0.555, Med: -0.041, Q3: 0.611&   & \\
		
		\quad $\beta_0$ (Intercept)     & 4.204 (0.211)          &  [3.788, 4.621]     \\
		\quad \texttt{religiousness}    (\texttt{No} vs. \texttt{Yes})  & -0.146 (0.135) & [-0.412, 0.120]   \\
		\quad \texttt{emotional$\_$stability} & -0.232 (0.029) &  [-0.289, 0.174]    \\
		\quad \texttt{university}   (\texttt{No} vs. \texttt{Yes})  & -0.280 (0.119)    & [-0.516, -0.044]  \\
		pseudo-$R^2=0.507$ &  &  \\
		&                   &          \\
		\textbf{Log-Normal Model}:    &  &  \\
		Residuals quantiles: Q1: -0.298, Med:-0.060, Q3: 2.427 & & \\
		
		\quad $\beta_0$ (Intercept)    & 0.138 (58.573)    &  [7.792,8.336]   \\
		\quad \texttt{religiousness}  (\texttt{No} vs. \texttt{Yes})  & 0.256 (0.088)  & [0.083,0.429]  \\
		\quad \texttt{emotional$\_$stability}  & 0.003 (0.019)  & [-0.034,0.041]  \\
		\quad \texttt{university}     (\texttt{No} vs. \texttt{Yes})   & 0.162 (0.078)  & [0.008,0.316] \\
		pseudo-$R^2=0.199$ &  &  \\
		&                   &                  \\
		\textbf{Fuzzy Normal Linear Model}:	&  &   \\
		Residuals quantiles: Q1: -0.287, Med:0.068, Q3: 0.737 & &  \\
		\quad $\beta_{0}$ (Intercept) & 3.383 (0.259) &  [2.870,3.894] \\ 
		\quad \texttt{religiousness} (\texttt{No} vs. \texttt{Yes})  & -0.169 (0.152) & [-0.469,0.130]\\ 
		\quad \texttt{emotional$\_$stability} & -0.127 (0.034) & [-0.195,-0.060] \\ 
		\quad \texttt{university} (\texttt{No} vs. \texttt{Yes}) & -0.172 (0.134) & [-0.436,0.093] \\ 
		pseudo-$R^2=0.304$ &  &  \\
		\bottomrule
	\end{tabular}
\caption{{Application}: Estimates, standard errors and CIs of the Normal linear model on crisp ratings, Log-Normal linear model on response times, and the fuzzy Normal linear model on the composite indicator \texttt{depression}. Note that the categorical variable have been codified with dummy coding with the following reference levels: \texttt{religiousness} (ref.: \texttt{No}), \texttt{university} (ref.: \texttt{No}). For all the analyses, $\alpha=0.05$.}
\label{tab2}	
\end{table}

\section{Conclusions}\label{sec4}

In this article we have provided initial findings to the problem of analyzing ratings data and response times with fuzzy numbers. This constitutes a crucial problem, especially when researchers need to evaluate the subjective component of questionnaire and survey data. A novel fuzzy rating procedure has been used (fIRTree), which allows for combining a probabilistic model of rater's uncertainty and response times in a unique fuzzy representation. The novelty of the solution lies in the fact that response times and ratings data can be integrated using a common formal representation, which is in turn easy to interpret and use. A real case study has been discussed to highlight the characteristics of the proposed approach and a proper fuzzy data analysis has been adopted. Unlike for standard data analyses on ratings and response times, the proposed procedure is more parsimonious with regard to the number of statistical analyses because it requires estimating a single statistical model instead of two separated models for crisp ratings and response times and the number of hypothesis testing procedures preserves the interpretation of the inferential results by using single linear tests on model's results. However, the fIRTree-based application has been focused on personality questionnaire data only, as they provide a more natural context for interpreting ratings data. Further studies might also include cognitive test-based data, which provide a well-suited framework for the joint modeling of responses and times \cite{de2019overview}. In a similar way, additional studies might also evaluate the extension of fuzzy linear models to cope with the non-negativity of fuzzy responses, such as gamma or ex-Gaussian fuzzy linear models. {Considering the mapping between empirical data (i.e., rating responses, response times) and fuzzy numbers, a fully fuzzy solution based on fuzzy inferential systems might be used in order to map fuzzy numbers and empirical data. For instance, this could potentially improve the overall interpretability of the proposed~method.}

\clearpage
\bibliographystyle{unsrt} 
\bibliography{biblio}

\clearpage

\end{document}